\documentstyle[12pt,aasms4,psfig,epsf]{article}
\received{00 }
\accepted{00 }
\def\ddeg   {\hbox{$.\!\!^\circ$}}              
\def\sec    {$^{\rm s}$}                        
\def\kms    {\hbox{km{\hskip0.1em}s$^{-1}$}}    
\def\i      {\hbox{\it I}}                      
\def\dasec  {\hbox{$.\!\!^{\prime\prime}$}}     
\def\dasec  {\hbox{$.\!\!^{\prime\prime}$}}     
\def\dsec   {\hbox{$.\!\!^{\rm s}$}}            

\def\sec    {$^{\rm s}$}                        
\def\kms    {\hbox{km{\hskip0.1em}s$^{-1}$}}    
\def\i      {\hbox{\it I}}                      
\def\etal   {{\it et al. }}                     

\centerline{Submitted to the Editor of the Astrophysical Journal {\it 
Letter} date }

\lefthead{Yusef-Zadeh et al}
\righthead{OH(1720 MHz) Masers Near the Galactic Center}

\def\kms{km s$^{-1}$}

\begin{document}

\title{Three New Supernova Remnant OH Masers Near the Galactic Center: Evidence 
for Large Scale  Maser Emission from Supernova Remnants}

\author{F. Yusef-Zadeh}
\affil{Department of Physics and Astronomy, Northwestern University, 
Evanston, Il. 60208 (zadeh@nwu.edu)}

\author{W. M. Goss}
\affil{National Radio Astronomy Observatory, P.O. Box 0, 
Socorro, New Mexico 87801 (mgoss@nrao.edu)}

\author{D. A. Roberts}
\affil{University of Illinois, 1002 Green St., Urbana, IL 61801
(dougr@ncsa.uiuc.edu)}

\author{B. Robinson}
\affil{Department of Physics and Astronomy, Northwestern University,
Evanston, Il. 60208}

\author{D. A. Frail}
\affil{National Radio Astronomy Observatory, P.O. Box 0, 
Socorro, New Mexico 87801 (dfrail@aoc.nrao.edu)}



\def\dasec  {${\rlap.}^{\prime\prime}$}    
\def\sec    {$^{\rm s}$}                        
\def\kms    {\hbox{km{\hskip0.1em}s$^{-1}$}}    
\def\i      {\hbox{\it I}}                      
\def\etal   {{\it et al. }}                     

\begin{abstract}

A survey of the inner 8$^\circ \times 1^\circ$ of the Galactic plane
toward the Galactic center has been carried out at the 1720 MHz
transition of OH molecule using the VLA in its D configuration
with a resolution of $\approx$70$''\times45''$. 
The detection of compact 1720 MHz OH masers associated with
three supernova remnants G357.7+0.3, G1.13--0.1 (Sgr D) and
G1.4--0.1 as well as  new extended maser line emission from
 G357.7+0.3 and G357.7--0.1 (the Tornado Nebula) were then followed up by 
A-array observations with  spectral and spatial resolutions 
of 0.3 \kms\ and $\approx$3$''\times2''$, respectively. 

The 1720 MHz OH maser line emission is considered to be a powerful
shock diagnostic and is collisionally pumped by H$_2$ molecules at
the site where C-type supernova shocks drive into adjacent
molecular clouds. The new observations show clear evidence of 
extended  features coincident  with compact and bright masers, 
the best example of which is 
a coherent  feature over a scale of about 20 pc surrounding 
the shell of the SNR G357.7+0.3. 
We argue that this remarkable feature 
is an OH maser and is physically associated with the remnant.  
This implies that 
the ambient molecular cloud is uniform in its
density and temperature with restricted range of pumping conditions
and survives the 
passage of a large-scale shock front.

\end{abstract}
\bigskip\noindent {\it Subject headings}:  
Galaxy: center ---ISM: molecules -- supernova remnants -- magnetic fields, 
masers -- shock waves

\section{Introduction}

A series of papers have recently examined the interaction site of
supernova remnants (SNR's) with molecular clouds, almost two decades after
the initial discovery of OH(1720 MHz) maser emission toward
supernova remnants (SNRs) by Goss \& Robinson (1968). These so-called
``supernova remnant masers'' (Fukui 1995) signal perhaps the best evidence
for the site of physical interaction between molecular clouds and 
SNRs. In particular, the masers signify   
shocked gas as the
expanding remnant runs into the  nearby molecular cloud (Elitzur 1976;
Lockett, Gauthier \& Elitzur 1998; Wardle, Yusef-Zadeh \& Geballe
1998). 
In addition, the study of OH(1720 MHz) masers in the Galactic center region have
provided the estimates of the line of sight magnetic field behind a shock
front, the size and shape of scatter-broadened sources masked by
the
turbulent screen, the length scale of the magnetic fluctuations in the
scattering medium and the magnitude of the tidal shear experienced by an
expanding nonthermal shell into a dense molecular cloud (Yusef-Zadeh et
al. 1999).

The  revival of OH(1720 MHz) maser study  
was initiated by the association of 26 OH(1720) 
masers  with W28, a well known SNR interacting with
an adjacent molecular cloud (Frail, Goss \& Slysh 1994). 
Following the survey 
of well-known SNRs in the disk of the Galaxy (Frail et al. 1996;
Green et al 1997; Koralesky et al. 1998), we searched for these masers in 
 the   central degree of the Galactic center where 
two nonthermal sources Sgr A East and G359.1--0.5 
showed association with OH(1720 MHz) masers (Yusef-Zadeh
et al. 1995, 1996).
 Using the  Very Large Array of the National Radio Astronomy 
Observatory\footnote{The National Radio Astronomy Observatory is a facility
of the National Science Foundation, operated under a cooperative agreement
by Associated Universities, Inc.} the 
present survey covers   an expanded area 
near the Galactic center; we report the detection of 
three new  SNR masers (G357.7+0.3, G1.05-0.15 and G1.4-0.1 as seen in 
Figure 1) within the inner few degrees of 
the Galactic center. We also report the detection of extended OH maser
emission associated with G357.7+0.3 and G357.7--0.1 (the Tornado Nebula). 
The latter source was known to have an associated compact OH maser emission 
(Frail et al. 1996).

\section{Observations}
 
The OH 1720 MHz maser line observations were made 
on 19 and 28 August 1996,
using the compact D configuration of the VLA.  A total of 
88  antenna pointings were used  covering roughly the inner 
8$^\circ \times 1^\circ$ ($l \times b$) of the Galactic center region. 
Figure 1 shows a sketch of 85  circles, each of which 
represents the 26$^\prime$ FWHM of the primary beam.  
 The filled circles in this figure show the positions where 
new OH(1720 MHz) masers   have 
 been detected. 
   Figure~1 also shows three additional  filled circles 
associated with  the Galactic center, Sgr A, G359.5--0.18 and  
W28; in these    cases supernova maser emission 
had been reported previously (Yusef-Zadeh, Uchida \& Roberts 1995; 
Frail, Goss \& Slysh 1994; Yusef-Zadeh et al. 1996, 1999).  Roughly
5  minutes were spent observing each field  using the four IF mode of the
VLA correlator in two overlapping spectral windows.  Each IF pair
simultaneously observed right and left circular polarization in 127
channels with a total bandwidth of 1.5625 MHz (273 \kms).  In order to
maximize a total velocity coverage, the two IF pairs were centered
at $V_{\rm LSR}$ = \hbox{$-$80} and +80 \kms, respectively.

The calibration of the absolute flux density, the complex gains, and
the bandpass response was carried out with the Astronomical Image
Processing System (AIPS) of the NRAO.  The data from the two IF pairs
covered  a total velocity
range  of $\pm$216 \kms.  Data from the
line-free channels were used to remove the continuum from the line
data in the visibility plane with the AIPS task UVLSF.  Final
imaging, image processing, visualization, and profile analysis were
carried out with the Multichannel Image Reconstruction, Image Analysis
and Display (MIRIAD) system of the Berkeley-Illinois-Maryland
Association (BIMA).  The images were made with natural weighting for
optimal sensitivity with  beam size of 
$\approx70'' \times 45''$ image 
and Hanning smoothed off-line to give a velocity
resolution of 4.25 \kms.  

Follow-up high-resolution observations were  carried out on
November 22 and December 24, 1996 in the A-array 
configuration of the VLA in
order
to improve the spatial and velocity structure of the masers
initially found in the low-resolution D array 
observations.  We observed
four
fields centered on G357.7+0.3 ($\alpha$, $\delta$[ 1950 ] =
$17^{\rm h} 35^{\rm m}$ 17\dsec7, $-30^\circ 32^\prime
01^{\prime\prime}$ at $V_{\rm LSR}$=--35 \kms), G357.7--0.1
($17^{\rm h} 36^{\rm m}$ 54\dsec3, $-30^\circ 56^\prime$ 15\dasec0
at $V_{\rm LSR}$=--5 \kms), Sgr D ($17^{\rm h} 45^{\rm m}$
43\dsec0, $-28^\circ 10^\prime$ 25\dasec0 at $V_{\rm LSR}$=--5
\kms) and G1.4--0.1 ($17^{\rm h} 46^{\rm m}$ 19\dsec2, $-27^\circ
46^\prime$ 52\dasec0 at $V_{\rm LSR}$=--3 \kms) in the 1720 MHz
hyperfine transition of OH molecule.  Using both the right ($RCP$)
and left ($LCP$) hands of circular polarization, the 195.3 kHz
bandwidth and 127 channels gave a velocity resolution of 0.27
\kms\ after online Hanning smoothing.  The final images had typical
synthesized beam of 2\dasec5 $\times$ 1\dasec3 (PA=3\ddeg9)
and typical rms noise of $\approx$12 mJy beam$^{-1}$ per channel.
Images of Stokes {\it V} and {\it I} based on A array observations 
were also made for the three new  
sources.  The upper limits to 
 their line-of-sight magnetic field via the Zeeman effect are measured.
The polarization data 
indicate 3-$\sigma$ upper limits  of 1.7, 2.2, 1.7, 0.6, 0.5  mG  for  G1.4--0.1,  
G357.7--0.1,  G357.7+0.3 OH1720:A1, A2, A5 (see Table 1).

\section{Results}

The D array survey showed three new sources of maser emission  associated 
with nonthermal radio continuum sources. A new extended OH(1720 MHz)
 emission is  also detected toward the Tornado nebula, as described below. 
Table 1 lists  the Gaussian fitted position and velocity of the compact 
sources as well as their deconvolved sizes based on A array  observations. 
The errors to the positions depend on signal-to-noise ratio.

\subsection{G357.7+0.3 (The Square Nebula)}

The continuum source G357.7+0.3 is a probable SNR 
(Reich \& F\"urst 1984; Leahy 1989; Gray 1994); 
the radio source is linearly polarized and has a steep spectral 
index, with a  shell-like 
appearance and its associated  X-ray emission. 
The structure of this unusual
continuum source is noted for its nearly square-like  morphology
(Gray 1994). There is also an appearance of distorted structure to
the NW of the continuum feature. Figure 2 shows the grayscale
continuum image of the source with a spatial resolution of
83$''\times43''$ 
based on Mologolo Observatory Synthesis Telescope
(MOST)  observations at 843 MHz made by Gray (1994).
Superposed on this image are naturally weighted contour image of 
maser emission averaged  between --42 and --29 \kms\ with a resolution
of $100''\times100''$ based on the D array data.
We note that the contours of OH(1720 MHz) emission mimic closely the
shape of the continuum structure to the NW of the continuum image.
In this region the extended OH(1720 MHz) structure is seen on a scale of 15$'$.
Since the phase center of the VLA observations was placed to the
NW of G357.7+0.3,
 (at $\alpha(1950)={\rm 17^h 35^m 17.8^s},
\delta(1950)=-30^\circ 32^\prime 01^{\prime\prime}$) the best
sensitivity to detecting the line emission lies in this region of
the remnant. The contours shown to the SW may have been
affected  by the primary beam correction, which amplifies the
noise at the edge of the primary beam. 
However, it is clear that the OH(1720 MHz) line feature 
is quite large, the   15$'$ extent 
outlines the eastern half  of this 
unusual  nonthermal   feature. 
The large-scale  OH feature is noted best at a velocity of
--35.4 \kms where an evelope of weak emission with a flux 
density of $\approx$20 mJy/beam is distributed on a 
 a scale of 15$'$; Several bright 
sources  on the scale of the beam which is 86$''\times45.7''$ 
are surrounded by the large scale OH feature. 
Most of the extended 
emission was resolved out in high resolution data 
implying that the emission is diffuse. The peak diffuse 
emission is  127 mJy/beam with the beam size of 
69$''\times34''$ (at $\alpha(1950)={\rm 17^h 35^m 17.78^s},
\delta(1950)=-30^\circ 31^\prime 21^{\prime\prime}$)
should have been detected in the A array images 
with a sensitivity of 52 mJy/beam 
if the emission were collection of point sources. 
It is noteworthy 
that the strongest maser emission near --35 \kms\ 
coincides with  the NW region
of the remnant where  the  continuum  distribution   
shows a prominant departure from circular-shaped geometry.  


Table 1 lists the peak positions and central 
velocities of five   
sources (A1 -- A5) seen in the A array 
image. 
The bright peak emission  in 
low-resolution data, as shown  in the contours of 
Figure 2,
breaks up into 
five  sources in the A array  observations.
The crosses on this figure represent the position of 
compact sources 
seen in the A-array observations.  
Although much of the 
extended features seen in Figure  2 are resolved 
out in the A array images,
the extended emission with the resolution of 
3\dasec2 $\times$ 1\dasec5
at a level of 
$\sim$ 5 to 25  mJy beam$^{-1}$ corresponding to 
T$_b \approx$ 530 to 2600 K
is  detected  between the 
compact and bright maser sources listed in Table 1.  We also note 
that source A3 has a size of $\approx4.9''\times2.5''$ 
with a position angle of 42$^0$. This position clearly shows that 
the compact features are also resolved along the 
direction of the  SNR shell.  
The   peak flux density of the compact sources is
$\approx$500 mJy beam$^{-1}$ which correspond to 
T$_b\approx 5\times10^4$  K. Two spectra  showing the 
extended emission observed both in the D and A array data are
shown 
in Figure 3. The D array emission spectrum (left) 
which arises from the position 
$\alpha(1950)={\rm 17^h 35^m 07^s},  \delta(1950)=-30^\circ 32^\prime
41^{\prime\prime}$ 
is not detected due to its large size of more than 2$'$ 
in the  A array data whereas  
the A-array spectrum (right) 
arises from a diffuse region of size 5$''\times7''$ 
near $\alpha(1950)={\rm 17^h 35^m 17.8^s},  \delta(1950)=-30^\circ 32^\prime
16.5^{\prime\prime}$. The extended  
components show similar kinematics as the 
compact components at  velocities near --35 \kms.

\subsection{G357.7--0.1 (the Tornado Nebula)}

This elongated radio continuum source called the Tornado
nebula is known to have a 
nonthermal origin with a spectral index consistent with 
a shell-type SNR estimated to be located at a distance greater than 
6 kpc (Radhakrishnan et al. 1972; 
Shaver et al. 1985a; Helfand \& Becker 1985;
Stewart, et al. 1994). 
The compact source at the western edge of G357.7--0.1 is 
considered to be an HII region based on its IRAS spectrum and 
$^{12}$CO emission at +9 \kms\ (Shaver et al. 1985b; see an
alternative
interpretation by Shull, Fesen \& Shaken 1989). Frail 
et al. (1996) detected an OH(1720 MHz) maser source at the 
western edge of the Tornado (beam of $\approx15''$) 
and to the north of 
the bright compact continuum source at a  velocity of --12.4 \kms\
with a peak flux density of 277 mJy.
  
The A array observations with a resolution of 
3\dasec2 $\times$ 1\dasec5
detect this  maser spot with a  position in good agreement with Frail et al. (1996).
Table 1 lists the properties of the resolved maser spot with a flux density
of 209 mJy at the velocity of --12.2 \kms. The low-resolution observations
show the compact maser spot as well as a newly detected extended OH(1720 MHz)
structure to the northeast of the nonthermal feature. The extended structure
is best seen in Figure 4 where the contours represent the  line
emission averaged  between --13.1 and --8.7 \kms\ with a spatial resolution
of 114$''\times38''$ based on the D array data.
 The high-resolution grayscale continuum image of the
Tornado nebula (Shaver et al. 1985a)
is also shown. 
The cross represents the 
position of the maser spot detected in A array 
observations. The  extended line
emitting feature 
is observed  at a velocity close to the velocity of the 
compact source coinciding  with
the brightest continuum feature. 
   The extended line 
emission is weak with a typical flux density ranging 
 between 25 and 40 mJy beam$^{-1}$
(rms noise  2.5 mJy beam$^{-1}$). The morphology and 
the kinematics of  OH(1720 MHz)  
features as well as their brightness  temperatures suggest that the
compact and extended structures
are masers and are associated with the nebula (see section 4).   

\subsection{ Sgr D \& G1.4--0.1}

Sgr D is a  bright radio continuum source near the 
 Galactic center.  Sgr D consists of two
adjacent radio continuum sources.
One of the sources G1.13--0.10 is an HII region coinciding with a star
forming region and the other is SNR G1.05--0.15 with nonthermal
characteristics which is considered to be located near or beyond the 
Galactic center 
(Downes et al. 1979; Odenwald \& Fazio 1984;
Liszt 1992; Gray 1994; Mehringer et al. 1998).

Figure 5 taken from the continuum survey by 
Liszt (1992),  shows a mosaic continuum image of the 
Sgr D region located in  the bottom half of the 
figure with a spatial resolution 
of $30''\times15''$ PA=0$^0$. The two bright continuum sources
G1.13--0.10 and G1.05--0.15
are observed  adjacent to each other
near 
$\alpha(1950)={\rm 17^h 45^m 20^s},  \delta(1950)=-28^\circ 05^\prime$.
The weak shell-like structure G1.4--0.1 
(to the north)  about 15$'$  NW  
of the Sgr D region  (Liszt 1992; Mehringer et al. 1998). 

Our  low-resolution OH survey observations 
 showed two compact maser spots toward G1.4--0.1 
and G1.05--0.15 at velocities of --2.5 and --1.4 \kms, respectively. 
The low-resolution  peak of 
G1.4--0.1 consists of two components whereas the 
maser spot in G1.05--0.15
is unresolved in the  A array observations.
These    compact OH(1720 MHz) maser spots 
associated with nonthermal sources are 
 represented by a crosses in Figure 4. 
The properties of these 
maser spots are  described  in Table 1. 
We also noted extended thermal emission toward 
the thermal component of the Sgr D region in the D array data. 
Contours 
in this figure represent OH(1720 MHz) emission at a velocity of
$-16.2$ \kms\ based on D array observations.  

\section{Discussion}

The common OH masers at 1665, 1667, and 1612 MHz are generally associated
with HII regions and evolved stars and are thought to be pumped by
far-IR radiation. 
The class of OH(1720 MHz) masers is distinguished from traditional OH 
masers by being spatially and kinematically associated with SNRs, 
shows none of the other  transitions of OH and  
is probably pumped collisionally. Recent
theoretical and observational studies suggest that the OH(1720 MHz)  masers
are associated with C-type shocks and are collisionally pumped in
molecular clouds at temperatures and densities 50 -- 125 K and $10^5-10^6$
cm$^{-3}$ (Lockett et al. 1999; Wardle et al. 1998; Frail and Mitchell
1998).  The evidence for C-type shocks is also suggested  by infrared
emission lines of H$_2$O and OH from a number of interacting SNRs with
molecular clouds (Reach and Rho 1998a, 1998b). In a recent model, Wardle et
al. (1998) argue that the X-ray photodissociation in SNR masers is
responsible for enhancing the OH abundance behind the C-type shock fronts.
The evidence for the association of a number of composite SNRs (i.e. center-filled
morphology in X-rays and shell-like morphology in radio continuum) and SNR
masers including G357.7+0.3 is consistent with this suggested model.

Another distinction that
may characterize  traditional OH masers and 
SNR OH masers is their linear sizes. 
 We note the presence of a remarkable extended
OH(1720 MHz) feature in  G357.7+0.3 distributed 
over an angular size of $15'$  
corresponding  to   27 pc at
the distance of 6.4 kpc (Leahy 1989). Such an extended 
coherent structure
which  has been observed  in at least three  other
 supernova maser sources (G359.1--0.5, 3C391 and Tornado),  is 
argued below to have maser characteristics. 
This association  
provides additional support to the large-scale nature  of the  
interaction of the
SNR with adjacent  molecular clouds. 

What is the nature of spatially extended OH(1720 MHz) 
emission? One possibility is that these 
prominent extended features 
are  due to slightly enhanced excitation temperature 
of thermal foreground objects having 
a similar origin to widespread Galactic OH(1720 MHz) 
emission observed in the Galactic plane using single 
dish telescopes (Haynes and Caswell 1977; Turner 1982). 
Alternatively, these features could be  low-gain masers and co-exist 
 with compact masers as has been argued 
in the case of  G359.1-0.5 (Yusef-Zadeh, Uchida and Roberts 1995). 

Based on the spectra shown in Figure 3, we believe that the 
emitting  extended OH features
observed  in both the D and A array images of 
G357.7+0.3 
are associated with the compact OH(1720 MHz) 
maser spots having  high brightness
temperatures. First, both the extended OH features
coincide with the 
radio continuum edges. However,  
both the compact and extended OH(1720 MHz) features  
show  no clear correlation with the strength of 
background  radio continuum emission
from  the SNRs. In the case of widely spread OH(1720 MHz) 
masers, strong background radio sources either from thermal 
or nonthermal continuum radio sources are 
required. The extended OH(1720 MHz) features show a range of 
brightness temperatures between 10 and 2500 K on 
scales varying from  several arcseconds to several arc minutes; 
no 1720 MHz absorption lines have been detected toward this
source. 
The brightness temperature of the extended 
emitting features at some locations is much greater than the 
gas temperature. Also,  the brightness temperature of the line emission is 
much less than the the brightness temperature of the 
continuum emission.  The lack of detecting any observed OH (1720 MHz) 
absorption lines 
argues against 
a thermal origin of  the optically thin extended OH features.
 In the  88 D array search positions 
in the Galactic center, 
 extended OH (1720 MHz) emission 
is detected in four regions with compact 1720 MHz maser spots.
This coincidence 
suggets that the extended and compact OH (1720 MHz) 
sources are physically associated. 

Second, the compact
source A3 in G357.7+0.3, (see Table 1, 
size  3.5$''$) is oriented along 
 the direction of the nonthermal shell in agreement
with the  large-scale extended OH structures. 
The largest size of the scatter-broadened OH(1720 MHz) 
maser sources associated with 
Sgr A East  at the Galactic center is only 1.3$''$
(Yusef-Zadeh et al. 1999).
It is unlikely that the size of source A3 can be 
explained by scatter  broadening; Unless G357.7+0.3 lies
behind an extended HII region, the resolved sizes of 
OH sources given in Table 1 are likely to be due to 
intrinsically  extended nature of OH(1720 MHz) 
maser emission.  
Furthermore, the narrow line widths of a few \kms\ 
for the  extended features 
are  similar  to  the compact sources; 
these line widths are an order of  magnitude 
less than the linewidth of 
molecular clouds in the Galactic center region (Bally \etal 1988).
The kinematics and the location of the compact and extended 
 features suggest that they are  located  at the boundary of 
a   --35 \kms\  molecular cloud  
and the G357.7+0.3  nonthermal shell source.

The remarkable similarity in the shape of 
the large-scale extended  OH(1720 MHz) emission  and the 
nonthermal  continuum distributions
implies  that the  system of SNR-molecular clouds
are  interacting. 
Another signature of the physical interaction between
SNRs and molecular clouds may be the highly distorted 
appearance of the remnants (e.g. G357.7+0.1 and 
the Tornado nebula G357.7--0.1).
 G357.7+0.3 has  a 
square-shape geometry  and  
the region to its NW appears to be most distorted (see the original 
figure by Gray 1994). 
There are no known molecular observations of the ambient gas in the 
vicinity of G357.7+0.3, the extended OH(1720 MHz)  emission 
may well suggest the existence of 
 an extended molecular cloud surrounding G357.7+0.3.
The  fact that the SNR deviates from 
a shell-like geometry 
at the interaction site may suggest the following picture: 
The pressure of 
X-ray gas in  the interior of the expanding shell  may not be 
sufficient  to dominate   the magnetic field pressure 
behind the shock front leading to a distortion of 
 the shell-like structure of the remnant as it collides with the dense 
molecular cloud.  
Future Zeeman measurements of the
extended OH(1720 MHz) masers as well as molecular 
line observations have the
potential to quantify the physical conditions of the 
shocked gas and the ambient cloud
as well as to determine the 
scale lengths over which  the 
magnetic field are organized behind the large-scale shocks.

It is  possible that large-scale OH(1720 MHz) maser emission is 
a distinct character of these types of masers. 
The cloud must be relatively homogeneous in its
 gas  temperature and 
density in order for OH(1720 MHz) maser to be pumped 
collisionally on such a scale under the restricted conditions.
 Unlike 
traditional OH masers, these large-scale OH(1720 MHz) masers
point to a global low-gain population inversion of OH in  the 
 extended ambient 
cloud caused by the passage of a shock front.
The 
present data are ambiguous 
in distinguishing  whether the variations of the
brightness temperature, as seen in Figure 2, results 
from  gas temperature  or  gas density variations.
The compact maser sources arise along the edge of the supernova shells
where the acceleration is considered to be
perpendicular to the line of sight and where the velocity coherence
is achieved with a small velocity gradient along the line of sight.
For large-scale  maser sources, the acceleration is unlikely to  be
perpendicular
to the line of sight throughout a latge region of the cloud, thus resulting  
a velocity difference between   the compact and extended sources (e.g.
3C391). 

In conclusion, we have searched for signs of 
physical interaction between nonthermal 
radio continuum sources 
and molecular clouds in the Galactic 
center region by using the OH(1720 MHz) maser emission 
as a probe. 
Three new masers   which are all associated with 
SNRs have been  discovered in this survey. 
These detections plus  three SNR masers from 
previous measurements  bring a total of 
six SNR masers found 
among 17 searched   SNRs (Green 1998)
toward  the Galactic center region.
Although the sample of searched SNRs is  incomplete, 
the high detection rate of 35\% in the  longitudes 
between --4.5$^0$
and 5.5$^0$ is  more than three
times the rate of detection found 
 elsewhere in the disk of the Galaxy (e.g. Green et al. 1997). 
 This high rate of detection 
may be explained by high density 
of molecular gas  distributed throughout the 
inner few degrees of the Galactic center, thus 
providing the physical conditions required  to 
produce OH(1720 MHz) emission. 

{\bf {Acknowledgments}}: We thank W. Cotton, A. Gray and C. Salter for
providing continuum images and Mark Wardle for useful discussions.

\begin{figure}
\centerline{\psfig{file=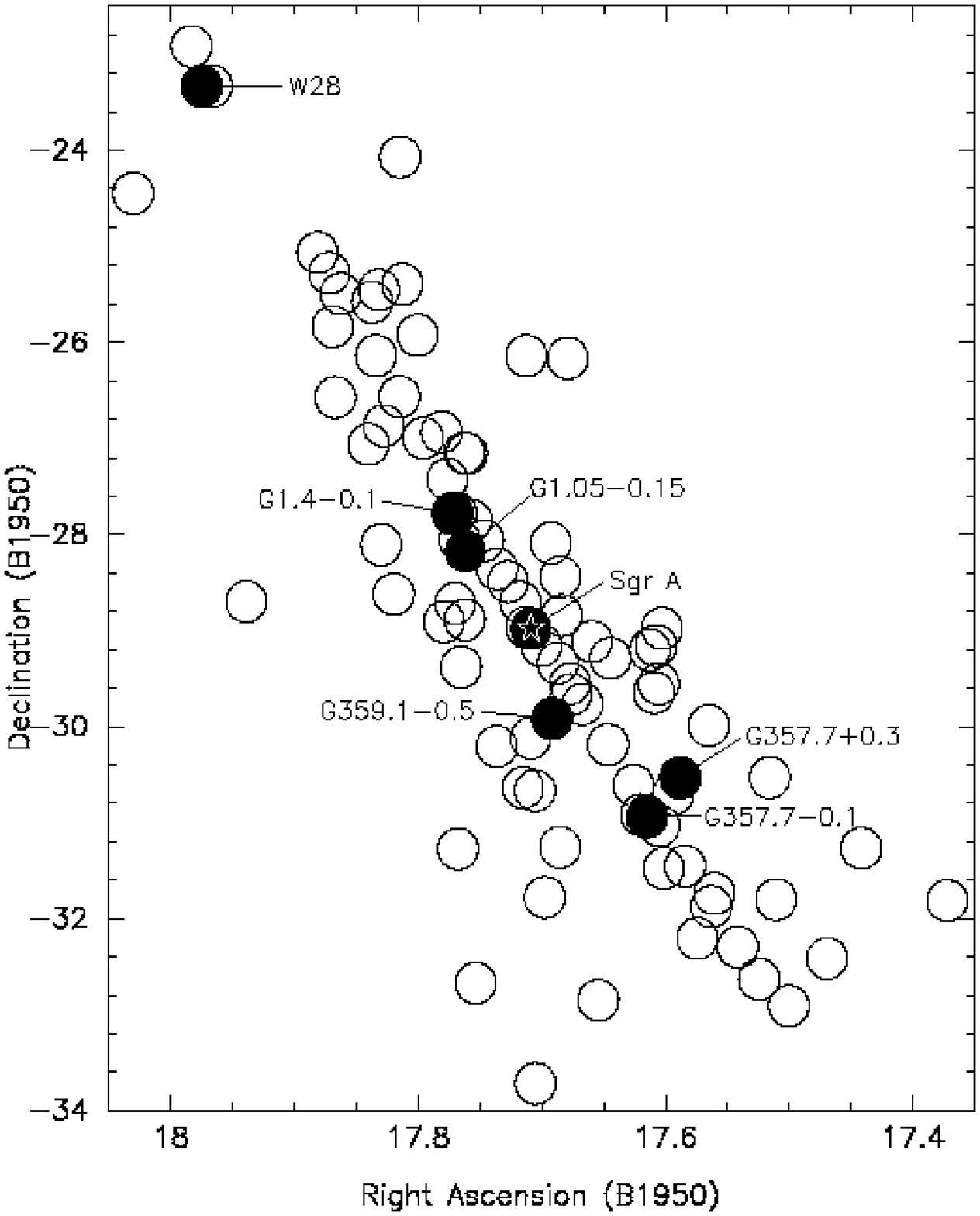,height=5.0in}}
\figcaption{ A representation of the area searched in the 88 pointings of
the D-array 1720 MHz OH observations along the Galactic plane 
near the Galactic center.  The size  of each circle (26$'$
diameter)
corresponds to 
the full width at half maximum of the primary beam at 1720 MHz. 
The filled circles coincide with regions where OH(1720 MHz)
masers associated with SNRs have been detected from this and previous 
observations. The star symbol corresponds to the position of 
Sgr A$^*$ at the Galactic center.}
\end{figure}

\begin{figure}
\centerline{\psfig{file=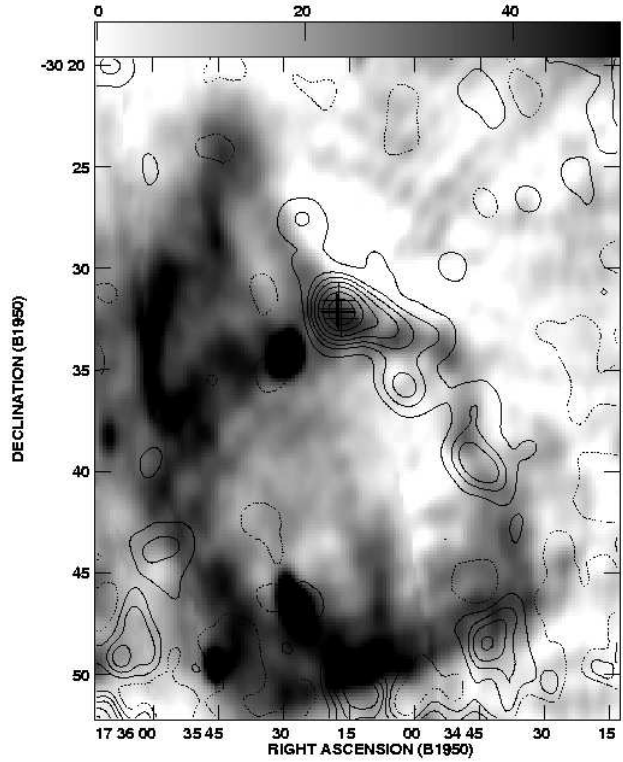,height=5.0in}}
\figcaption{ A grey-scale continuum image of G357.7+0.3 
shown in black   with a
resolution of $83''\times43''$ (Gray 1994) superimposed 
on the contours of
velocity averaged  OH(1720 MHz) emission between
--42 and --29 \kms\ convolved with beam of 
$100''\times100''$ having an rms noise of 3.2 mJy beam$^{-1}$. 
The levels are  32.5 $\times$ (--3, 3, 8, 
13, 20, 30, 45, 65) mJy beam$^{-1}$ \kms. The 
crosses at the peak contour position corresponds to the 
position of maser spots  detected in the A array 
observations with a resolution of 
2\dasec8 $\times$ 1\dasec4. }
\end{figure}

\begin{figure}
\centerline{\psfig{file=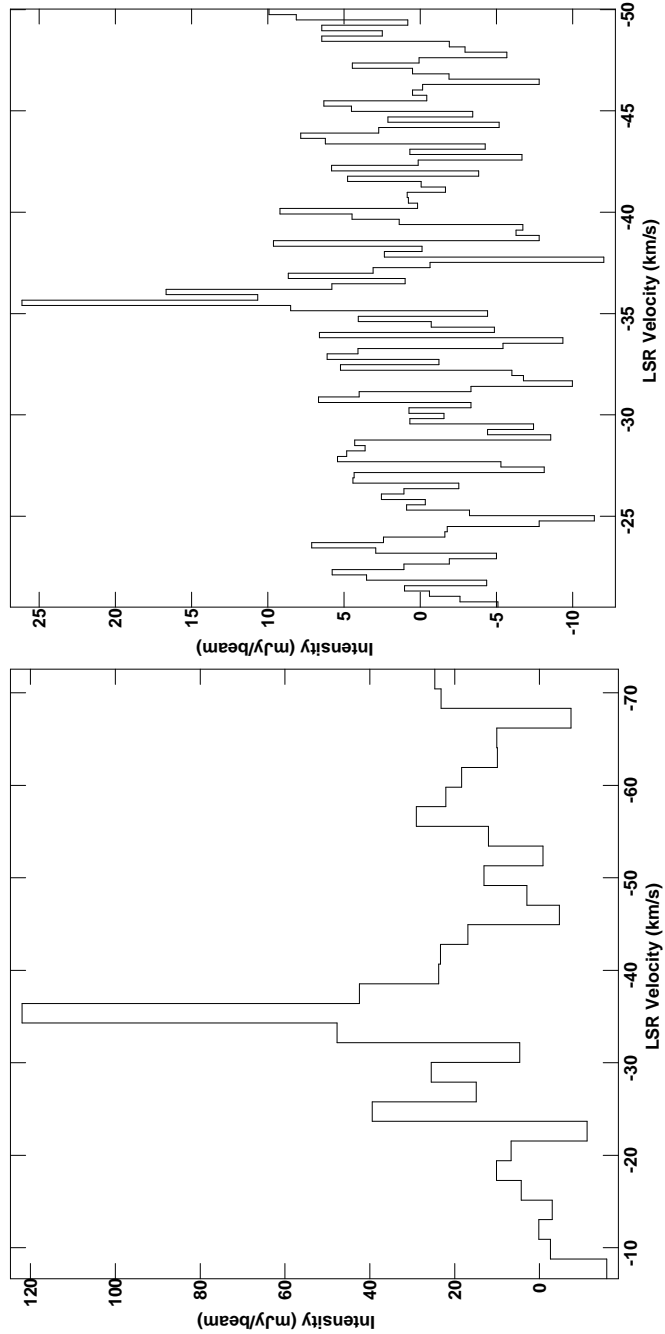,height=7.0in}}
\figcaption{ Two Stokes $I$ spectra from the diffuse regions of 
G357.7+0.3 based on the D array data (left) 
with a spatial resolution of 
$69'' \times 34''$ 
and the A array data (right) with a resolution of 
2\dasec8 $\times$ 1\dasec4.
The left panel  centered at  
$\alpha(1950)={\rm 17^h 35^m 07^s},  \delta(1950)=-30^\circ 32^\prime
41^{\prime\prime}$ whereas the right spectrum is integrated 
over a 5$'' \times  7''$ region centered on 
$\alpha(1950)={\rm 17^h 35^m 17.8^s},  \delta(1950)=-30^\circ 32^\prime
16.5^{\prime\prime}$}.
 
\end{figure}

\begin{figure}
\centerline{\psfig{file=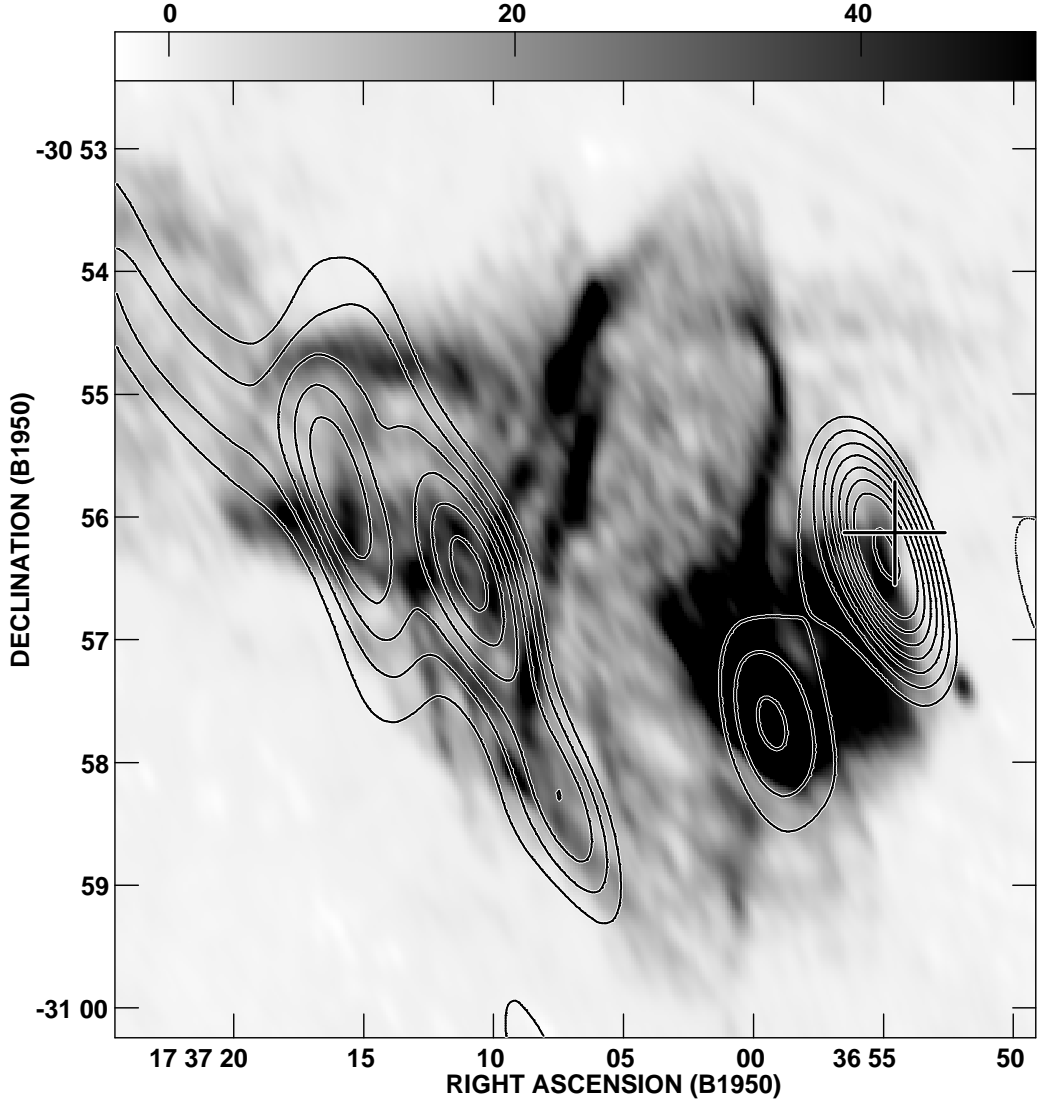,height=7.0in}}
\figcaption{ A grey-scale continuum image of G357.7--0.1 shown in black
with  a
resolution of $14.6''\times4.9''$ provided by C. Salter 
with contours of velocity averaged  OH(1720 MHz) emission between 
--13.1 and --8.7 \kms\ at  a
spatial resolution of 
$114''\times38''$ (PA=20$^0$) superposed. The levels are  22 $\times$ (--3, 3, 4, 
5, 6, 7, 8, 9, 11, 13, 15)  mJy beam$^{-1} \kms$. The 
cross   corresponds to the 
position of the  1720 MHz maser spot observed  with a resolution of 
3\dasec2 $\times$ 1\dasec5. }
\end{figure}

\begin{figure}
\centerline{\psfig{file=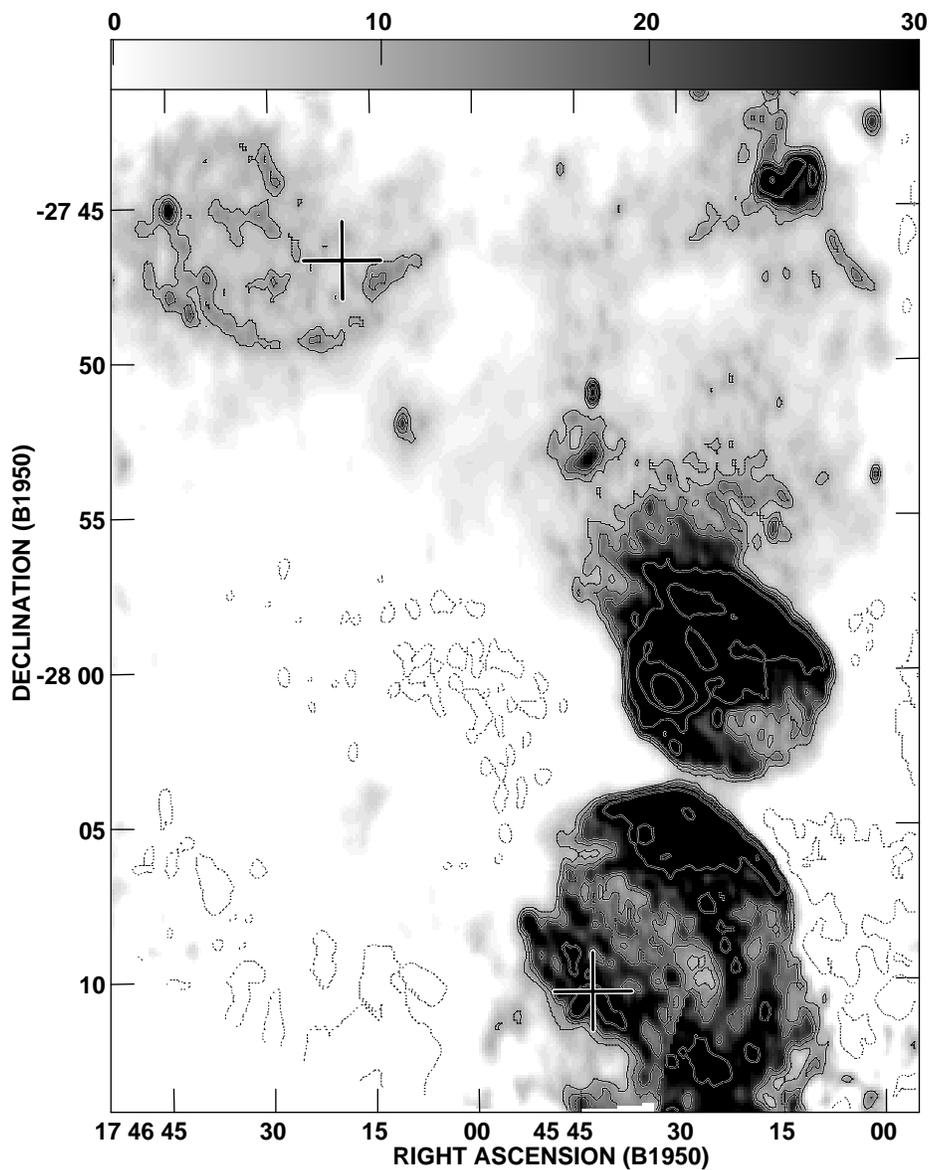,height=7.0in}}
\figcaption{ A grey-scale continuum mosaic  image
displayed in black shows
  prominent thermal  source G1.13--0.1 to SW and the 
nonthermal sources 
G1.05--0.15 and G1.4--0.1 
with a resolution of
30$''\times15''$ (Liszt 1992) near 
$\alpha, \delta(1950)={\rm 17^h 45^m 30^s}, -28^\circ 08^\prime$
and ${\rm 17^h 46^m 35^s},  -27^\circ 47^\prime$, 
respectively.
  Contours of thermal OH(1720 MHz)
emission associated with G1.13--0.1 at a velocity of --16.2 \kms\ 
are superposed.  The contour levels are 10 $\times$ (--3, 3, 4, 5, 6, 7, 8, 9, 
11, 13 and 15) mJy beam$^{-1}$. 
The crosses  represent  the 
positions of OH(1720 MHz) maser spots with a resolution of
2\dasec75 $\times$ 1\dasec35 and an rms noise per channel of 
21 mJy beam$^{-1}$. }
\end{figure}

\begin{table}
\centerline{\psfig{file=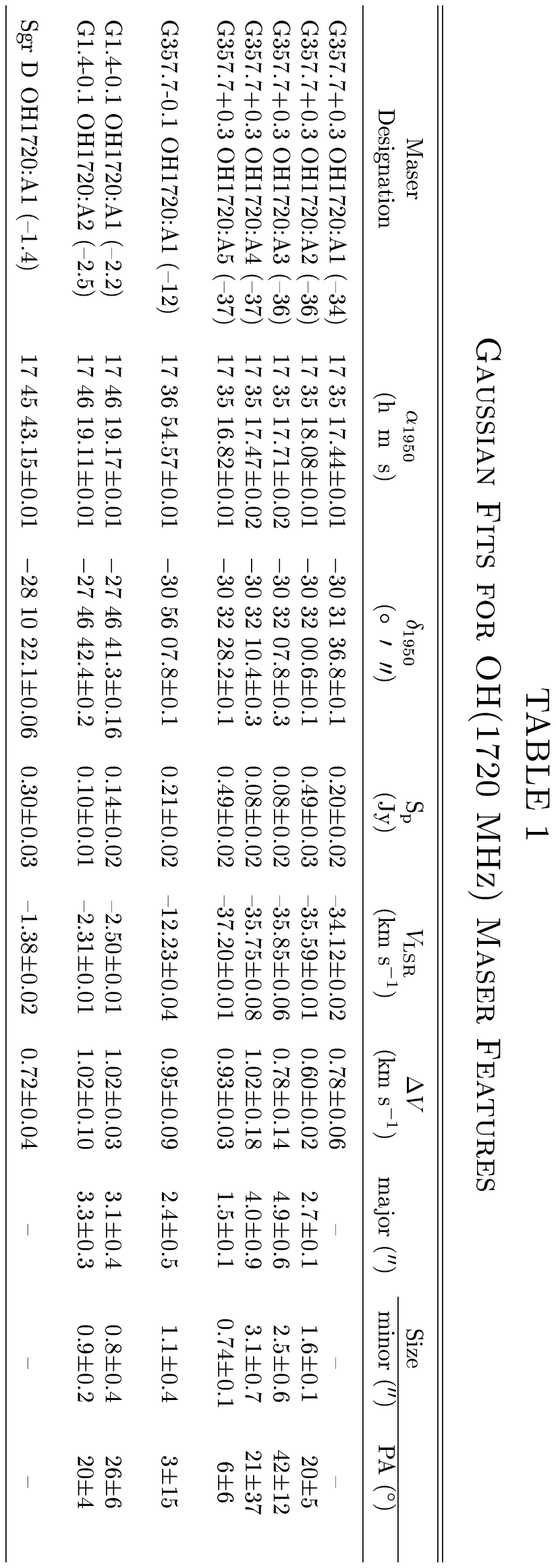,height=8.0in}}
\figcaption{ Table 1}
\end{table}


\begin{references}

\reference{b98} Bally, J., Stark, A.A., Wilson, R.W., \& Henkel, C. 1988, 
ApJ,  324, 223


\reference{c97} Claussen, M.J., Frail, D.A., Goss, W.M., \& Gaume,
R.A. 1997, ApJ 489, 143

\reference{e76} Downes, D., Wilson, T., Bieging, J., \& Wink, O.
1980, A\&AS, 35, 1 

\reference{e76} Elitzur, M. 1976, ApJ, 203, 124

\reference{e76} Frail, D.A. \& Mitchell, G.F. 1998, ApJ, 508, 690.


\reference{f94} Frail, D.A., Diamond, P.J., Cordes, J.M. \& van
  Langevelde, H.J. 1994, ApJ, 427, L43

\reference{f96} Frail, D. A., Goss, W. M., Reynoso, E. M., 
Giacani, E.B., Green, A. J. \& Otrupcek, R. 1996, AJ, 111, 1651

\reference{fgs94} Frail, D.A., Goss, M.W. \& Slysh, V.I. 1994, ApJ,
  424, L111

\reference{f95} Fukui, Y. 1995,
  Science 270, 1771

\reference{g68} Goss, W.M. \& Robinson, B.J. 1968, ApJ, 2, L81

\reference{g94} Gray, A.D. 1994, MNRAS, 270, 835

\reference{g98} Green, D.A. 1998, A catalogue of Galactic Supernova Remnants
(1998 September version), Mullard Radio Astronomy Observatory 
Cambridge, UK (http://www.mrao.cam.ac.uk/surveys/snrs/) 

\reference{g114} Green A.J., Frail, D.A., Goss, W.M. \& Otrupcek, R.
 1997, AJ, 114, 2058

\reference{hc77} Haynes, R.F.  \& Caswell J.L.   1977, 
MNRAS, 178, 219 


\reference{hb85} Helfand, D.J. \& Backer, R.H.  1985, Nature, 
313, 118

\reference{k98} Koralesky, B., Frail, D.A., Goss, W.M., 
Claussen, M.J. \& Green, A.J. 1998, AJ, 116, 1323
 

\reference{l89} Leahy, D.A. 1989, ApJ, 216, 193


\reference{l92} Liszt, H.S. 1992, ApJS, 82, 495


\reference{l98} Lockett, P., Gauthier, E. \&  Elitzur, M. 1998, 
ApJ, 511, 235

\reference{m98} Mehringer, D.M., Goss, W.M., Lis, D.C., Palmer, 
P. \& Menten, K.M. 1998, ApJ, 493, 274

\reference{of84} Odenwald, S.F. \& Fazio, G.G, 1984, ApJ, 283, 601

\reference{rr98a} Reach, W.T. \& Rho, J. 1998a, ApJ, 507, L93

\reference{rr98b} Reach, W.T. \& Rho, J. 1998b, ApJ, 511, 836

\reference{r72} Radhakrishnan, V., Goss, W.M., 
Murray, J.D.  \& Brooks, J.W.  1972, ApJS, 24, 49


\reference{rf84} Reich, W. \& F\"urst, E. 1984, A\&AS, 57, 165

\reference{s85a} Shaver, P.A., Salter, C.J., 
Patnaik, A.R., van Gorkom, J.H. \& Hunt, G.C. 1985a, Nature, 313, 113

\reference{s85b} Shaver, P.A., Pottash, S.R., Salter, C.J., 
Patnaik, A.R., van Gorkom, J.H. \& Hunt, G.C. 1985b, A\&A, 147, L23

\reference{s94} Stewart, R.T., Haynes, R.F. 
Gray, A.D. \& Reich, W. 1994, ApJ, L39

\reference{w98} Wardle, M., Yusef-Zadeh, F., \& Geballe 1998, 
preprint (astro-ph 9811090).

\reference{y95} Yusef-Zadeh, F., Uchida, K.I., \& Roberts, D.A. 1995,
  Science 270, 1801

\reference{y96} Yusef-Zadeh, F., Roberts, D.A., Goss, W.M., Frail, D.A. \&
  Green, A. 1996, ApJ 466, L25

\reference{y99} Yusef-Zadeh, F., Roberts, D.A., Goss, W.M., Frail, D.A. \&
  Green, A. 1999, ApJ, 512, 230.

\end{references}
\end{document}